\documentclass[reprint,superscriptaddress,amsmath,amssymb,aps,prl,floatfix]{revtex4-2}

\usepackage{graphicx}
\DeclareGraphicsExtensions{.pdf,.png,.eps}

\usepackage{dcolumn}
\usepackage{bm}
\usepackage[hyperfootnotes=false]{hyperref}
\hypersetup{
    colorlinks=true,
    linkcolor=red,
    filecolor=magenta,      
    citecolor=blue
}

\usepackage[utf8]{inputenc}

\makeatletter
\def\paragraph#1{%
   \textit{#1}%
  \unskip 
}
\makeatother

\newcommand{\alphay}{\alpha_Y}
\newcommand{\lamy}{\lambda_Y}

\newcommand{\GN}{G_{\mathrm{N}}}

\begin{document}
	
\title{Fifth-Force Constraints from UV-Complete Scalar-Tensor Gravity}

    \author{Alfio M. Bonanno}
    \email{alfio.bonanno@inaf.it}
	\affiliation{INAF -- Osservatorio Astrofisico di Catania, Via S.\ Sofia 78, 95123 Catania, Italy}
	\affiliation{INFN, Sezione di Catania, Via S.\ Sofia 64, 95123 Catania, Italy}
	
	\author{Emiliano M. Glaviano}
    \email{emiliano.glaviano@inaf.it}
	\affiliation{INAF -- Osservatorio Astrofisico di Catania, Via S.\ Sofia 78, 95123 Catania, Italy}
	\affiliation{INFN, Sezione di Catania, Via S.\ Sofia 64, 95123 Catania, Italy}
    \affiliation{Dipartimento di Fisica e Astronomia, Università di Catania, Via S. Sofia 64, 95123, Catania, Italy}

\date{\today}

\begin{abstract}
\noindent We study an O$(N)$ scalar multiplet nonminimally coupled to gravity and follow its renormalization-group (RG) flow in the vicinity of an interacting, nonperturbatively UV-complete scaling regime of scalar-tensor theory. In the broken phase, the radial mode mediates a universal Yukawa correction to Newtonian gravity, parametrized by a strength $\alphay$ and range $\lamy$. Imposing UV completeness—regular RG trajectories that reach the UV scaling regime—restricts the infrared data to a finite wedge, which maps to a narrow region in the $(\alphay,\lamy)$ plane. Its complement is, therefore, ruled out by UV completeness alone. Remarkably, part of this theory-excluded domain lies below current experimental exclusion envelopes, so improved fifth-force searches can directly test and potentially falsify this class of UV-complete scalar-tensor models.
\end{abstract}

\maketitle

\paragraph{Introduction—}Yukawa-type deviations from Newtonian gravity provide a universal parametrization of hypothetical fifth forces \cite{Fischbach:1996eq,Fischbachbook,PhysRevLett.56.3} and are the target of a broad experimental program spanning laboratory, geophysical, and astrophysical tests (see, e.g., Ref.~\cite{Will:2014kxa} for a review). In this context, a generic modification of the Newtonian potential takes the form
\begin{equation}
	V(r) = -\frac{\GN m_1 m_2}{r}\left[1 + \alphay\,e^{-r/\lamy}\right]
	\label{eq:Yukawa-potential}
\end{equation}
and is constrained with high precision by torsion-balance experiments \cite{Kapner:2006si,Adelberger:2009zz}, geophysical surveys \cite{Anderson1989}, lunar laser ranging \cite{Williams:2004qba}, and planetary ephemerides \cite{Fienga2011}. Fifth forces arise naturally in theories with light scalar fields---including moduli and dilatons \cite{Damour:1994zq,FujiiMaedaBook}---and thus constitute a sensitive probe of physics beyond the standard model.
	
A question that has received comparatively little attention is whether \emph{quantum gravity itself} restricts the allowed values of $(\alphay,\lamy)$. In phenomenological analyses these parameters are treated as arbitrary infrared quantities to be bounded experimentally. However, if gravity and matter admit a nonperturbatively UV-complete description, the renormalization group (RG) flow may not allow every infrared choice of couplings. Determining which fifth-force parameters survive the requirement of UV completeness is,  therefore, a qualitatively new theoretical criterion.
	
Asymptotic safety provides a concrete framework in which this question can be addressed. scalar-tensor extensions of gravity \cite{Quiros:2019ktw,Zhang:2023nil,Huang:2024gvi} are known to possess a non-Gaussian fixed point where both the Newton coupling and the scalar nongaussian coupling approach finite interacting values \cite{Percacci:2009fh,Narain:2009fy,Hamada:2017rrp}. A striking feature of this fixed point is that scale invariance is intrinsically quantum: the RG flow selects a nonclassical nongaussian coupling rather than the classical conformal value, and this behavior persists in functional RG studies of the full scalar dependence \cite{Percacci:2015wwa,bonanno2025propertimefunctionalrenormalizationon}.

In this work, we determine which Yukawa parameters $(\alphay,\lamy)$ are compatible with scalar-tensor theories whose RG trajectories remain regular and reach the ultraviolet scaling regime. We find that UV-complete trajectories populate only a narrow, finite region of fifth-force parameter space, while its complement is excluded by UV completeness alone. Remarkably, part of this theory-excluded domain lies below existing experimental bounds, so forthcoming short-range and Solar System searches can directly test and potentially falsify this class of UV-complete scalar-tensor models.

It is instructive to place our analysis in the broader context of quantum-gravity consistency conditions that have been discussed in recent years, in particular swampland-inspired criteria \cite{ArkaniHamed:2006dz,Palti:2019pca}. A representative example is the weak gravity conjecture, which has been invoked to argue that extremely weak long-range forces may be incompatible with certain classes of ultraviolet completions. The ultraviolet consistency requirement implemented here is of a different type: it is defined by the existence of regular renormalization-group trajectories approaching an interacting scaling regime of scalar-tensor gravity. In this sense, our results provide a complementary notion of quantum-gravity consistency. Comparing the resulting constraints with swampland expectations offers an additional perspective on how quantum gravity may restrict low-energy modifications of gravity.

\paragraph{Nonperturbatively UV-complete scalar-tensor model—}We consider an O$(N)$ multiplet $\mathbf{\Phi}= (\phi_1,...,\phi_N)$, which transforms as a fundamental representation of the O$(N)$ group, nonminimally coupled to gravity via the Euclidean action
\begin{equation}
S = \int d^d x\,\sqrt{g}\left[
-F(\rho)\,R
+ \frac12 \sum_{i=1}^N\phi_i(-\Box)\phi_i
+ U(\rho)
\right],
\label{eq:action}
\end{equation}
with $\rho = \frac12 \sum_{i=1}^N \phi_i^2$ and matter fields minimally coupled to $g_{\mu\nu}$.
We introduce dimensionless variables,
\begin{equation}
x = k^{2-d}\rho,\quad
u(x,t) = k^{-d} U(\rho),\quad
f(x,t) = k^{2-d} F(\rho),
\end{equation}
with RG time $t = \ln(k/k_0)$.  Here $k$ denotes the functional renormalization-group coarse-graining scale, implying that fluctuations with characteristic momenta $p^2 \lesssim k^2$ are suppressed; its interpretation as a resolution scale is discussed below. 

In the spontaneously broken phase, the O$(N)$ multiplet can be parametrized as $\mathbf{\Phi}_a=\phi_1 n_a+\phi_a$ with $n_a=\delta_{a1}$ and $\mathbf{n}\cdot\phi=0$. A vacuum $\rho_0$ selects a direction in field space, $\phi_1 = v + \sigma$, $\phi_a = \pi_a$ $(a>1)$, and $\rho = \rho_0 + \mathcal{O}(\sigma,\pi^2)$. Since $F$ and $U$ depend only on $\rho$, only the radial mode $\sigma$ couples linearly to $R$ and mediates a fifth force,
while the $N-1$ transverse modes $\pi_a$ remain derivatively coupled.

The RG flow of scalar-tensor systems of the form in Eq.  (\ref{eq:action}) has been investigated extensively. Early studies mostly relied on finite-dimensional truncations for $F(\rho)$ and $U(\rho)$ \cite{Percacci:2003jz,Narain:2009fy,Narain:2009gb,Merzlikin:2017zan}.

A key structural result, however, is that ultraviolet completion is governed by an interacting \emph{global scaling regime} of the \emph{dimensionless} functions $f(x,t)$ and $u(x,t)$ themselves:
in the broken phase, the functional flow admits solutions for which
\begin{equation}
f(x,t)\to f_\ast(x),\qquad u(x,t)\to u_\ast(x)\qquad (k\to\infty),
\end{equation}
i.e.\ the fixed-point conditions $\partial_t f=\partial_t u=0$ hold for all $x$. Such scaling solutions are found without imposing an \emph{a priori} ansatz for the $x$-dependence and have been constructed explicitly within Wetterich-type flows \cite{Percacci:2015wwa,Labus:2015ska}. The same scaling regime is reproduced within the proper time formulation of the functional RG \cite{Bonanno:2019ukb,bonanno2025propertimefunctionalrenormalizationon}, supporting its robustness against changes in flow equation and coarse graining.  We therefore use a local \emph{parametrization} of the running functions near the minimum to isolate the couplings relevant for fifth-force phenomenology.  Concretely, we parametrize $u(x,t)$ and $f(x,t)$ locally around the running minimum $x_0(t)$ as
\begin{equation}\begin{split}
&f(x,t) = \frac{1}{g(t)} + f_1(t)\,[x-x_0(t)],\\
&u(x,t) = u_0(t) + \lambda_4(t)[x-x_0(t)]^2,
\label{eq:truncation}
\end{split}\end{equation}
so that the relevant field derivatives of $f$ and $u$ are encoded at $x=x_0(t)$. The gravitational sector is encoded in the dimensionless Newton coupling $g(t)$ and the nongaussian coupling $f_1(t)$. 

In our analysis, we employ the proper time functional renormalization-group flow for gravity, first introduced in \cite{Bonanno:2004sy}, in which the cutoff is defined by a one-parameter family of regulators labeled by an integer $m$ ($m>2$  in $d=4$  dimensions). The parameter $m$ controls the localization of the momentum integration around the coarse-graining scale $k$. As discussed in \cite{Bonanno:2000yp} and more recently in \cite{Bonanno:2019ukb}, this class of regulators is known to yield highly accurate critical exponents, e.g. at the Wilson-Fisher fixed point.

In $d=4$, the induced flow of the couplings $(x_0,\lambda_4,g,f_1)$ admits an interacting fixed point in the broken phase,
\begin{equation}\begin{split}
&x_{0\ast}=\frac{N-7+\sqrt{3(11+N)}}{32\pi^2(m-1)}, \qquad\ \lambda_{4\ast}=0,\\
&g_\ast=\frac{192\pi^2(m-1)}{16-N},\qquad\;
f_{1\ast}=\frac{1}{6}-\frac{1-\sqrt{3\left(11+N\right)}}{6\left(N-1\right)}\,,
\end{split}\end{equation}
so that within Eq. (\ref{eq:truncation}), the fixed point exists for $1<N<16$. 

As expected for a finite-dimensional parametrization, the fixed-point coordinates $x_{0\ast}$ and $g_\ast$ are scheme dependent. They provide a local representation of the underlying global scaling regime that exists at the level of the dimensionless functions $f_\ast(x)$ and $u_\ast(x)$. By contrast, the fixed-point value of the nonminimal coupling $f_{1\ast}$ is found to be independent of the regulator and cutoff scheme within the class of proper time flows considered here. This robustness indicates that $f_{1\ast}$ is directly tied to the underlying functional scaling solution rather than to details of the truncation.

UV completeness further requires that RG trajectories, starting from $t=t_0$, remain regular and approach this scaling regime as $t\to\infty$. For fixed $g_0\equiv g(t_0)$,  $x_{0,0}\equiv x_0(t_0)$, and $f_{1,0} \equiv f_1(t_0)$, this imposes an upper bound on the quartic coupling at $t_0$, $\lambda_{4,0}<\lambda_{4,\rm cr}(f_{1,0};g_0)$, defining the UV-complete wedge in the IR data. In the large-$f_{1,0}$ regime relevant to the present analysis, $\lambda_{\rm cr}$ admits an asymptotic expansion,
\begin{equation}\begin{split}
\lambda_{4,\mathrm{cr}}
&(f_{1,0}\gg1)=\frac{8\pi^2}{3(N-1)\left(N-7+\sqrt{3}\sqrt{11+N}\right)}\frac{1}{f_{1,0}}
\\
&
-\frac{4\pi^2\left(2+N+2\sqrt{3}\sqrt{11+N}\right)}{9(N-1)\left(N-7+\sqrt{3}\sqrt{11+N}\right)^2}\frac{1}{f_{1,0}^2}
+ \mathcal{O}\!\left(\frac{1}{f_{1,0}^3}\right),
\label{eq:lambda-critical}
\end{split}\end{equation}
that is independent of $g_0$, $x_{0,0}$ and the regulator parameter $m$. The cutoff dependence remains very weak also outside the asymptotic regime, as illustrated in Table~\ref{tabm}.

      \begin{figure}[t]
    \centering
        \includegraphics[width=0.45\textwidth]{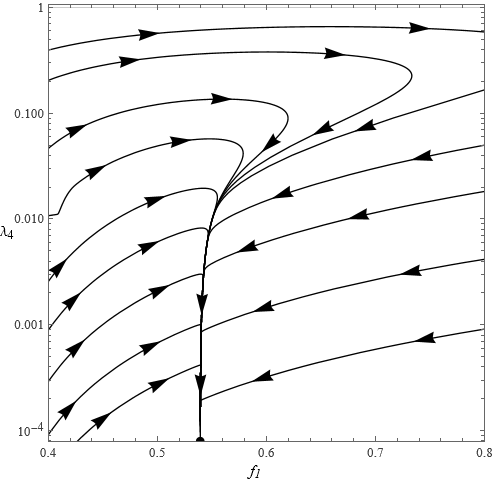}
        \caption{Renormalization-group flow in the $(f_1,\lambda_4)$ plane near the non-Gaussian scalar-tensor fixed point. Arrows indicate increasing RG time $t=\ln(k/k_0)$ with $k_0=1\,\mathrm{eV}$. Trajectories from a wide range of infrared initial conditions are attracted toward the fixed point, illustrating its UV-attractive character in the broken phase and the finite basin of attraction underlying the UV-complete wedge. The trajectories use initial data inside the wedge with $g_0=k_0^2/m_p^2=10^{-56}$ and $x_{0,0}=v^2/(2k_0^2)=1$.}
    \label{plotphase}%
    \end{figure}

\begin{table}
\centering
\begin{tabular}{c|ccc}
\hline
$m$ & $\lambda_{4,\mathrm{cr}}(5)$ & $\lambda_{4,\mathrm{cr}}(10)$ & $\lambda_{4,\mathrm{cr}}(50)$ \\
\hline
3  & 0.0677 & 0.0559 & 0.0240 \\
6  & 0.0674 & 0.0558 & 0.0237 \\
8  & 0.0678 & 0.0557 & 0.0243 \\
12 & 0.0673 & 0.0558 & 0.0242 \\
\hline
\end{tabular}
\caption{Values of $\lambda_{4,\mathrm{cr}}(f_{1,0})$ for representative values of $f_{1,0}$ and regulator parameter $m$, computed at fixed $g_0=10^{-56}$ and $x_{0,0}=1$. The variation with $m$ is very mild, indicating weak regulator dependence of the critical coupling.}
\label{tabm}
\end{table}

\paragraph{Phase dependence and breaking of marginality—}UV-complete trajectories arise \emph{only after} the flow enters the broken phase. In the symmetric phase the nongaussian coupling $f_1$ is exactly marginal at the non-Gaussian fixed point, as in earlier scalar-tensor analyses, so reaching the UV scaling regime requires a tuned infrared choice of $f_1$, and generic trajectories miss the UV critical surface.

Once the potential develops a nonzero minimum, $x_0(t)>0$, $\beta_{f_1}$ receives additional contributions depending on the location of the minimum and the curvature of the potential at $x_0$. These terms lift the exact marginality of $f_1$ and make it weakly relevant, producing an effective focusing: a finite-volume set of infrared initial conditions is funneled toward the UV scaling regime. This behavior is illustrated in Fig.~\ref{plotphase}.

Consistently, the stability matrix in the coupling space $(\lambda_4,x_0,g,f_1)$ at the interacting fixed point exhibits one exactly marginal eigendirection (within the present truncation), while the remaining directions are UV-attractive. Since the eigenvectors are generically linear combinations of couplings, the associated scaling fields are not aligned with the coordinate axes.
For $N=4$, the critical exponents read
\begin{equation}
\theta_i=\left(7-\frac{8\sqrt{5}}{3}\pm\frac{1}{3}\sqrt{1391-618\sqrt{5}},\;2,\;0\right).
\end{equation}
Inspection of the corresponding eigenvectors shows that the $\theta=2$ direction is essentially aligned with the Newton coupling $g$. The marginal and near-marginal eigendirections are also dominated by the gravitational sector but with subleading admixtures of $\lambda_4$ and $f_1$, rather than corresponding to a pure shift of the running minimum $x_0$.

\paragraph{Infrared matching scale—}In the functional RG, the scale $k$ is a coarse-graining parameter rather than a physical momentum transfer. To extract the Yukawa correction from the effective field equations, we couple the scalar-tensor sector to a classical macroscopic source, modeled as a perfect fluid. This description is meaningful only after coarse-graining over the microscopic constituents of matter (electrons and ions in ordinary media), i.e.\ on length scales larger than the characteristic microscopic response scales of the medium. It is, therefore, natural to define the phenomenological parameters $(\alpha_Y,\lambda_Y)$ at an infrared matching scale $k_0$ of order typical atomic-collective energy scales (eV range), where the continuum approximation for the matter sector is justified. For $k\lesssim k_0$, the flow is expected to effectively freeze: the microscopic constituents of the source carry masses and excitation gaps that suppress their quantum fluctuations in the RG propagators, so the matter sector is accurately described as a classical medium, and further running is strongly suppressed. In practice, we find that the running of the relevant dimensionless couplings has already become very mild in this regime, so that varying $k_0$ within a reasonable range around $k_0\simeq 1\,{\rm eV}$ does not change the qualitative structure of the UV-complete wedge in the $(\alpha_Y,\lambda_Y)$ plane \footnote{Numerically, $k_0\simeq 1\,\mathrm{eV}$ corresponds to a coarse-graining length $\ell_0\sim (\hbar c)/k_0 \approx 200\,\mathrm{nm}$, well above microscopic interatomic scales, and therefore, appropriate for a continuum, perfect-fluid description of matter.}.

\paragraph{Mapping to fifth-force parameters—}We express the phenomenology of the radial mode in terms of the Yukawa parameters $(\alphay,\lamy)$ in Eq.~(\ref{eq:Yukawa-potential}). Working in the Jordan frame, where matter couples minimally to $g_{\mu\nu}$, test bodies follow Jordan-frame geodesics, and the scalar-mediated force is universal (composition independent); nongaussianity resides in the gravitational sector through the term $F(\rho)R$ in Eq.~(\ref{eq:action}).

In the weak-field, static limit around the vacuum, the Yukawa strength and mediator mass are fixed by the scalar-tensor couplings evaluated at the running minimum $x_0(t)$. Adopting the standard weak-field reduction of scalar-tensor theories with a potential (see, e.g., \cite{Zhang:2023nil}), one obtains in our dimensionless variables
\begin{equation}\begin{split}
\alpha(t)
&=\left.\frac{2x_0\,f_x^2}{f+6x_0\,f_x^2}\right|_{x=x_0},
\qquad
\frac{m_s^2}{k^2}=\alpha(t)\,m_0^2 \equiv \tilde m_s^2,\\
m_0^2
&=\left.\frac{f\,u_{xx}}{f_x^2}\right|_{x=x_0},
\end{split}\label{eq:alpha-ms-general}
\end{equation}
where $f_x=\partial f/\partial x$ and $u_{xx}=\partial^2 u/\partial x^2$.

For the parametrization in Eq.~(\ref{eq:truncation}), this reduces to
\begin{equation}\begin{split}
&f(x_0)=\frac{1}{g(t)},\qquad f_x(x_0)=f_{1}(t),\\
&u_x(x_0)=0,\qquad u_{xx}(x_0)=2\lambda_{4}(t),
\end{split}\end{equation}
and therefore, at $t=t_0$,
\begin{equation}
\alpha_{Y}\equiv \alpha(t_0)
=\frac{2x_{0,0} f_{1,0}^2\,g_0}{1+6x_{0,0} f_{1,0}^2\,g_0},
\quad
\tilde m_s^{2}(t_0)
=\frac{4x_{0,0} \lambda_{4,0}}{1+6x_{0,0} f_{1,0}^2\,g_0}.
\label{eq:alpha-ms-trunc}
\end{equation}
Here, at fixed initial conditions $g_0$ and $x_{0,0}$, the values of $\lambda_{4,0}$ and $f_{1,0}$ are constrained to lie inside the UV-complete wedge according to Eq. (\ref{eq:lambda-critical}) and should not be confused with the ultraviolet fixed-point coordinates. The physical scalar mass at $t=t_0$ is $m_s(t_0)=\tilde m_s(t_0)\,k_0$, and the Yukawa range is $\lambda_{Y} \equiv 1/m_s(t_0)$. Equations~(\ref{eq:lambda-critical}) and (\ref{eq:alpha-ms-trunc}) thus provide a map from the RG initial data $(x_{0,0},\lambda_{4,0},g_0,f_{1,0})$ to the phenomenological parameters $(\alpha_{Y},\lambda_{Y})$ defined in Eq. (\ref{eq:Yukawa-potential}).

Combining this map with the UV-completeness condition, Eq.~(\ref{eq:lambda-critical}), implies that only a restricted region in the $(\alphay,\lamy)$ plane can be realized by UV-complete trajectories within our setup. Its complement is a theory-excluded domain: no choice of IR couplings compatible with the UV scaling regime can populate it. In particular, the very weak regulator dependence of $\lambda_{4,\mathrm{cr}}$ in Eq.~(\ref{eq:lambda-critical}) implies that the resulting exclusion bounds are, likewise, only very weakly scheme dependent, making the predictions robust under a change in the cutoff parameter $m$.


    \begin{figure*}[t]
    \centering
        \includegraphics[width=\textwidth]{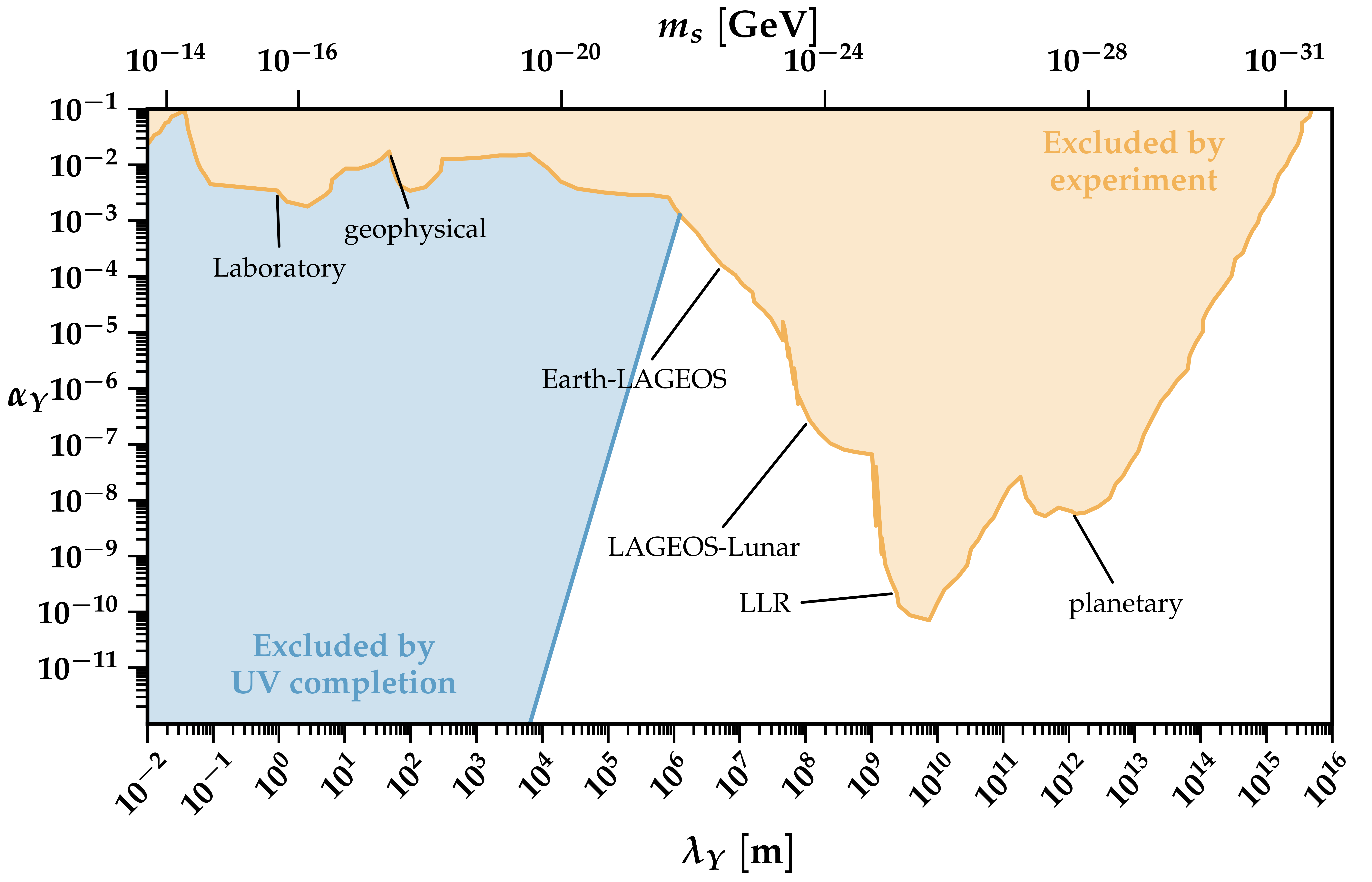}
        \caption{Constraints on Yukawa-type fifth forces in the $(\alpha_Y,\lambda_Y)$ plane. The orange region is excluded by existing laboratory, geophysical, and Solar System experiments, while the blue region is ruled out by the requirement of non-perturbative UV completeness of scalar-tensor gravity. The upper axis shows the corresponding scalar mass, $m_s$. Notably, part of the UV-excluded region lies below current experimental bounds, implying that future fifth-force searches can directly test this class of models. The plot is shown for $N=4$; the UV-excluded boundary depends only weakly on $N$ and is numerically very close for $N\neq4$, making the corresponding curves nearly indistinguishable on the scale of the figure.
        }
        \label{plotcomexp}%
    \end{figure*}

\paragraph{Comparison with existing bounds—}Laboratory, geophysical, and Solar System experiments constrain Yukawa interactions by providing 95\% C.L. upper limits on $\alphay(\lamy)$ over many decades in the force range $\lamy$. We adopt the published exclusion envelopes as conservative bounds,
\begin{equation}
\alphay(\lamy)\le \alpha_{\rm exp}^{95\%}(\lamy),
\end{equation}
without reanalyzing the underlying datasets or performing a global likelihood combination. The resulting comparison between experimental constraints and the RG prediction is shown in Fig.~\ref{plotcomexp} for $N=4$; the experimental exclusion envelopes are taken from the compilation in Ref.~\cite{Adelberger:2009zz}.

As visible in Fig.~\ref{plotcomexp}, a portion of the parameter space excluded by UV completeness lies below the current experimental envelopes. In this domain, existing tests would still allow a measurable Yukawa deviation from Newtonian gravity, while UV completeness forbids it. Improved torsion-balance, short-range, and precision Solar System probes targeting this region can,  therefore, directly falsify this class of UV-complete scalar-tensor theories.

\paragraph{Conclusions—}We have shown that non-perturbative UV completeness in scalar-tensor gravity constrains fifth-force phenomenology: only a restricted region in the $(\alphay,\lamy)$ plane is compatible with RG trajectories that remain regular in the broken phase and approach the interacting scaling regime in the UV. This restriction reflects the fixed-point structure and its phase dependence, which are found to be robust across different functional RG implementations of scalar-tensor systems.

While the precise boundary of the UV-allowed region is quantitative and can shift under the inclusion of higher-order operators, additional matter fields, or alternative parametrizations, the qualitative outcome persists: UV-complete trajectories populate only a limited subset of Yukawa parameter space. Extensions of this work include incorporating fermions and gauge fields, investigating possible screening mechanisms, and performing a refined statistical comparison with experimental datasets. Overall, our results illustrate how nonperturbative UV completion can translate into sharp, falsifiable predictions for low-energy tests of the inverse-square law.


\paragraph{Acknowledgments—}The authors thank Holger Gies for stimulating discussions, the Physics Department of the University of Jena for hospitality, and Chulalongkorn University for hospitality while part of this work was being completed.

\paragraph{Data availability—}The data that support the findings  of this article are not publicly available. The  data are available  from the  authors upon reasonable request.
    
\bibliographystyle{apsrev4-2}
%

\end{document}